\RequirePackage{fix-cm}
\documentclass[twocolumn,epjc3]{svjour3}  

\smartqed  % flush right qed marks, e.g. at end of proof
\RequirePackage{graphicx}

\usepackage{hyperref}
\usepackage{color}
\usepackage{amsmath}
\usepackage{subcaption}
\usepackage{lineno}
\usepackage{comment}
\usepackage{lmodern}
\usepackage{booktabs}
\usepackage{multirow}
\journalname{Eur. Phys. J. C}
\begin{document}
\newcommand{\nuc}[2]{$^{#2}\rm #1$}

\newcommand{\bb}[1]{$\rm #1\nu \beta \beta$}
\newcommand{\bbm}[1]{$\rm #1\nu \beta^- \beta^-$}
\newcommand{\bbp}[1]{$\rm #1\nu \beta^+ \beta^+$}
\newcommand{\bbe}[1]{$\rm #1\nu \epsilon \epsilon$}
\newcommand{\bbep}[1]{$\rm #1\nu \rm EC \beta^+$}

\newcommand{\pic}[5]{
       \begin{figure}[ht]
       \begin{center}
       \includegraphics[width=#2\textwidth, keepaspectratio, #3]{#1}
       \end{center}
       \caption{#5}
       \label{#4}
       \end{figure}
}

\newcommand{\apic}[5]{
       \begin{figure}[H]
       \begin{center}
       \includegraphics[width=#2\textwidth, keepaspectratio, #3]{#1}
       \end{center}
       \caption{#5}
       \label{#4}
       \end{figure}
}

\newcommand{\sapic}[5]{
       \begin{figure}[P]
       \begin{center}
       \includegraphics[width=#2\textwidth, keepaspectratio, #3]{#1}
       \end{center}
       \caption{#5}
       \label{#4}
       \end{figure}
}

\newcommand{\picwrap}[9]{
       \begin{wrapfigure}{#5}{#6}
       \vspace{#7}
       \begin{center}
       \includegraphics[width=#2\textwidth, keepaspectratio, #3]{#1}
       \end{center}
       \caption{#9}
       \label{#4}
       \vspace{#8}
       \end{wrapfigure}
}

\newcommand{\baseT}[2]{\mbox{$#1\times10^{#2}$}}
\newcommand{\baseTsolo}[1]{$10^{#1}$}
\newcommand{\THL}{$T_{\nicefrac{1}{2}}$}

\newcommand{\UBI}{$\rm cts/(kg \cdot yr \cdot keV)$}

\newcommand{\Uflux}{$\rm m^{-2} s^{-1}$}
\newcommand{\Ucpd}{$\rm cts/(kg \cdot d)$}
\newcommand{\Uexpo}{$\rm kg \cdot d$}

\newcommand{\Qbb}{$\rm Q_{\beta\beta}\ $}

\newcommand{\validate}{\textcolor{blue}{\textit{(validate!!!)}}}

\newcommand{\improve}{\textcolor{blue}{\textit{(improve!!!)}}}

\newcommand{\missing}[1]{\textcolor{red}{\textbf{...!!!...} #1}\ }

\newcommand{\missref}{\textcolor{red}{[reference!!!]}\ }

\newcommand{\quanta}{\textcolor{red}{\textit{(quantitativ?) }}}

\newcommand{\misscite}{\textcolor{red}{[citation!!!]}}

%K42
\newcommand{\PC}{$N_{\rm peak}$}
\newcommand{\BIC}{$N_{\rm BI}$}
\newcommand{\PAPR}{$R_{\rm p/>p}$}

\newcommand{\PCR}{$R_{\rm peak}$}

%Pd

\newcommand{\gline}{$\gamma$-line}
\newcommand{\glines}{$\gamma$-lines}
\newcommand{\gray}{$\gamma$-ray}
\newcommand{\grays}{$\gamma$-rays}

%general

\newcommand{\tab}{Tab.~}
\newcommand{\eq}{Eq.~}
\newcommand{\fig}{Fig.~}
\renewcommand{\sec}{Sec.~}
\newcommand{\chap}{Chap.~}

 \newcommand{\fn}{\iffalse \fi} %footnote explaination
 \newcommand{\tx}{\iffalse \fi} %text explaination
 \newcommand{\txe}{\iffalse \fi} %text extended explaination
 \newcommand{\sr}{\iffalse \fi} %section reference explaination
% \renewcommand{\cr}{\iffalse \fi} %citation reference explaination

%\title{Pushing the frontier in the search for rare decays of Gd isotopes}
%\title{Extending the Search for Rare Gd Isotope Decays}
\title{Improved sensitivity in the search for rare decays of Gd isotopes}


\author{
        B. Lehnert\thanksref{e1,addr2} %etc.
        \and
        S. S. Nagorny\thanksref{e2,addr5,addr1} %etc.
        \and
        M. Thiesse\thanksref{e3,addr3} %etc.
        \and
        F. Ferella\thanksref{addr1} %etc.
        \and    
        M. Laubenstein\thanksref{addr1} %etc.
        \and
        E. Meehan\thanksref{addr4}
        \and
        S. Nisi\thanksref{addr1}
        \and
        P. R. Scovell\thanksref{addr4}  %etc.
}

\thankstext{e1}{e-mail: bjoern.lehnert@tu-dresden.de}
\thankstext{e2}{e-mail: serge.nagorny@gssi.it}
\thankstext{e3}{e-mail: m.thiesse@sheffield.ac.uk}
% \thankstext{e4}{e-mail: paul.scovell@stfc.ac.uk}
% \thankstext{e5}{e-mail: emma.meehan@stfc.ac.uk}


\institute{Technische Universit\"at Dresden, Institut f\"ur Kern und Teilchenphysik, Zellescher Weg 19, Dresden, 01069, Germany \label{addr2}
           \and
           Gran Sasso Science Institute, viale F. Crispi 7, I-67100 L'Aquila, Italy\label{addr5}
           \and
           School of Mathematical and Physical Sciences, The University of Sheffield, Hounsfield Road, Sheffield, S3 7RH, U.K. \label{addr3}
           \and
           INFN - Laboratori Nazionali del Gran Sasso, Via G. Acitelli 22, 67100 Assergi (AQ), Italy \label{addr1}
           \and
           STFC, Boulby Underground Laboratory, Boulby Mine, Redcar-and-Cleveland, TS13 4UZ, U.K. \label{addr4}
}

\date{Received: date / Accepted: date}

% add and check reference
% check detector setup
%
%
%
%
%

\maketitle

\begin{abstract}

Gadolinium is widely used in multiple low-background experiments, making its isotopes accessible for rare decay searches both in-situ and through radiopurity screening data. This study presents an improved search for rare alpha and double-beta decay modes in $^{152}$Gd, $^{154}$Gd, and $^{160}$Gd isotopes using ultra-low background HPGe detectors at the Boulby Underground Screening (BUGS) facility. A total exposure of 6.7~kg$\cdot$yr of natural gadolinium was achieved using gadolinium sulfate octahydrate $(\text{Gd}_2(\text{SO}_4)_3 \cdot 8\text{H}_2\text{O})$ samples, originally screened for radiopurity prior to their deployment in the Super-Kamiokande neutrino experiment. Due to the detection methodology, only decays into excited states accompanied by gamma-ray emission were accessible. 
A Bayesian analysis incorporating prior experimental results was employed, leading to new lower half-life limits in the range of $10^{19} - 10^{21}$ years - an improvement of approximately two orders of magnitude over previous constraints. No statistically significant decay signals were observed. These results demonstrate the effectiveness of repurposing large-scale radiopurity screening campaigns for fundamental physics research.

\end{abstract}

%%%%%%%%%%%%%%%%%%%%%%%%%%%%%%%%%%%%%%%%%%%%%%%%%%%%%%%%%%%%%
%%%%%%%%%%%%%%%%%%%%%%%%%%%%%%%%%%%%%%%%%%%%%%%%%%%%%%%%%%%%%
%%%%%%%%%%%%%%%%%%%%%%%%%%%%%%%%%%%%%%%%%%%%%%%%%%%%%%%%%%%%%
\section{Introduction}
\label{intro}

Neutrinos are key particles in astrophysics. Although they were theoretically predicted in 1930 by W. Pauli~\cite{brown78} and experimentally detected for the first time in 1956~\cite{cowan56}, a comprehensive understanding of their properties remains elusive. The most significant unresolved issues concern the neutrino's mass and its nature~\cite{dolinski19,mohapatra07}. Among the various nuclear and electroweak processes involving neutrinos, neutrinoless double-beta decay offers the only practical means of obtaining definitive answers to both of these questions.

Neutrinoless double-beta decay is a hypothetical second-order nuclear process that is forbidden within the Standard Model of particle physics. However, if observed, it would establish the electron neutrino as a Majorana particle, implying that the neutrino possesses a non-zero effective Majorana mass term~\cite{Takasugi84,Schechter82,Avignone08}. This conclusion arises because this process is only possible if the neutrino is its own antiparticle. Furthermore, the existence of Majorana-type neutrinos could enable mechanisms that would explain the observed matter-antimatter asymmetry in the Universe~\cite{Fukugita86,Deppisch18}. Consequently, extensive worldwide experimental efforts are dedicated to this fundamental topic, with numerous international collaborations searching for neutrinoless double-beta decay in various isotopes~\cite{dolinski19,Pritychenko25}.

The current experimental sensitivity in this field has reached a half-life value of $10^{26}$ years, yet the process remains undetected. Consequently, next-generation experiments aim to increase their sensitivity by employing an extended active detector mass in the range of 1--10 tons and by reducing the background level below  $10^{-4}-10^{-5}$~counts/keV$\cdot$kg$\cdot$yr~\cite{Barabash23}. To achieve this exceptionally low internal background counting rate, all potential background components must be meticulously controlled and mitigated. Therefore, all future ultra-low-background (ULB) experiments will necessitate complex assays of all materials and reagents used, which must adhere to stringent radiopurity protocols specific to each individual experiment. Typically, such screening campaigns incorporate a combination of low-background measurements of material samples using high-purity germanium (HPGe) $\gamma$-spectrometers located deep underground, analysis of radon emanation, and assays using ICP-MS analysis (such as \cite{akerib20,aprile22}).

%On the other hand, these thorough material selections and studies of their radiopurity not only allow collaborations to significantly enhance experimental sensitivity and reach their main scientific goals, but also to investigate some rare nuclear processes as a by-product with an unprecedented sensitivity level. One notable example in this respect is the study of rare decays of natural Gd isotopes contained in Gd compounds used for neutron tagging and veto. Thermal neutrons, produced by high-energy cosmic muons in the atmosphere, surrounding rocks, detector construction materials, or through ($\alpha$,n) reactions on U/Th nuclides from internal contamination, are one of the most harmful background components \cite{}. Therefore, future ULB experiments are considering installing and using massive and effective neutron shields or neutron veto detectors \cite{}. Gadolinium, the element with the highest cross-section for thermal neutron capture, is often considered and used in veto and neutron tagging systems \cite{}.

In a different area of low-background, high-energy physics, the large-scale neutrino oscillation experiment, Super Kamiokande (SK), has dissolved 40 tons of a gadolinium salt (gadolinium sulfate octahydrate, Gd$_2$(SO$_4$)$_3\cdot8$H$_2$O) into their ultra-pure water Cherenkov detector to expand their scientific programmes and distinguish neutrinos from antineutrinos originating from supernova explosions. Consequently, before loading 13 tons of Gd$_2$(SO$_4$)$_3\cdot$8H$_2$O into SK in 2020~\cite{firstgdloading} and an additional 26 tons in 2022~\cite{secondgdloading}, samples representing a portion of the gadolinium salt underwent radiopurity screening at Boulby Underground Laboratory (UK). The raw material was required to be substantially radiopure to maintain SK's sensitivity to solar neutrinos and the diffuse supernova neutrino background (DSNB) signals. The requirements of $<5$~mBq/kg early-chain $^{238}$U, $<0.5$~mBq/kg late-chain $^{238}$U, $<0.05$~mBq/kg $^{232}$Th, and $<30$~mBq/kg $^{235}$U were largely achieved, although some batch-to-batch variation occurred~\cite{gdproduction}. The experimental data acquired within this radioassay campaign provides a unique opportunity to search for rare nuclear processes that may occur in natural gadolinium isotopes at an unprecedented sensitivity level.

%As shown in a recent study \cite{Gd2023}, under optimal experimental conditions, one could enhance the previously achieved sensitivity by several orders of magnitude by increasing the Gd sample mass and placing it in an optimized geometry on HPGe gamma spectrometers.

\begin{figure*}
  \centering
  \includegraphics[width=0.99\textwidth]{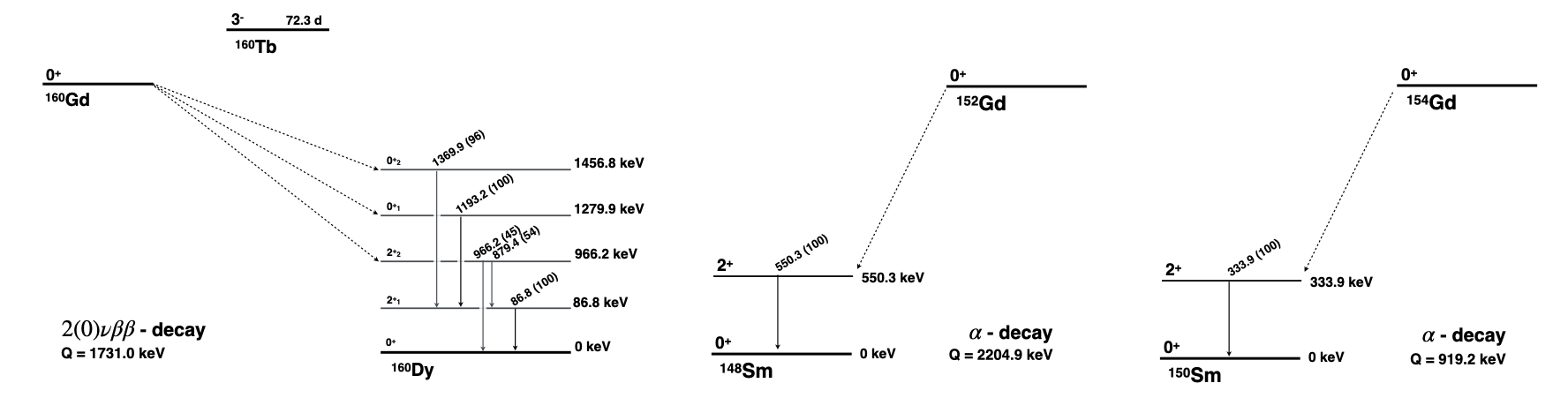}
  %\vspace{5cm}
\caption{Decay schemes for investigated decay modes in this work.}
  \label{pic:DecayScheme}
\end{figure*}

Therefore, this manuscript presents a new search for $\alpha$ and 2$\beta$ decay of Gd isotopes to excited levels of their daughter nuclides using radioassay data with a total exposure of 6.7 kg$\cdot$yr of natural Gd. Improved experimental sensitivity compared to the recent study~\cite{Laubenstein23} was achieved by using high-purity Gd-containing samples and increasing the sample exposure. However, as this experimental approach is best suited to studying rare decay modes with the emission of relatively high-energy de-excitation $\gamma$-rays (above 100 keV), the 2EC process in $^{152}$Gd is not considered here. Instead, the focus is on $\alpha$ and 2$\beta$ decay of \nuc{Gd}{152}, \nuc{Gd}{154}, and \nuc{Gd}{160} to excited states of their respective daughters. The decay schemes of all considered decay modes are shown in Figure \ref{pic:DecayScheme}.

%%%%%%%%%%%%%%%%%%%%%%%%%%%%%%%%%%%%%%%%%%%%%%%%%%%%%%%%%%%%%
%%%%%%%%%%%%%%%%%%%%%%%%%%%%%%%%%%%%%%%%%%%%%%%%%%%%%%%%%%%%%
%%%%%%%%%%%%%%%%%%%%%%%%%%%%%%%%%%%%%%%%%%%%%%%%%%%%%%%%%%%%%
%\clearpage
%\newpage
\section{Experimental Setup and Sample}

%Prior to the loading of 13 tons of Gd$_2$(SO$_4$)$_3\cdot$8H$_2$O to SK in 2020~\cite{firstgdloading} and an additional 26 tons in 2022~\cite{secondgdloading}, samples representing a portion of the gadolinium were screened for radiopurity at Boulby Underground Laboratory. The raw material was required to be significantly radiopure so that sensitivity to solar neutrinos and the diffuse supernova neutrino background (DSNB) signals in SK could be maintained. The requirements of \textless5 mBq/kg early-chain $^{238}$U, \textless0.5 mBq/kg late-chain $^{238}$U, \textless0.05 mBq/kg $^{232}$Th, and \textless30 mBq/kg $^{235}$U were largely achieved, though some batch-to-batch variation occurred~\cite{gdproduction}. 

The Gd$_2$(SO$_4$)$_3\cdot$8H$_2$O raw material was produced by the Nippon Yttrium Company in approximately half-ton batches between 2019 and 2022~\cite{firstgdloading,secondgdloading,gdproduction}. Two distinct production methods were employed, differentiated by their treatment of elemental radium. The second production method aimed to reduce radium concentration and thereby improve the late-chain thorium radiopurity. To verify the radiopurity of the produced batches, samples were sent to three underground screening laboratories worldwide, including the Boulby UnderGround Screening (BUGS) laboratory.

Following batch production, the manufacturer initially packed the samples in EVOH bags, each containing 5~kg ($\pm 0.5\%$) of material. These were shipped by aeroplane from the manufacturer in Japan to England. Upon arrival at the University of Sheffield, the samples were packed into type 448-G Marinelli beakers and transported to Boulby Underground Laboratory for HPGe screening.

Each batch of Gd$_2$(SO$_4$)$_3\cdot$8H$_2$O contained excess water from the production process. The amount of extra water in each batch was monitored by SK during Gd loading, averaging $4.4 \pm 1.0 \%$ w/w across the samples included in this analysis. Therefore, the mass of each sample in this study is taken to be $4.78 \pm 0.06$~kg of Gd$_2$(SO$_4$)$_3\cdot$8H$_2$O.

This study utilised 30 individual sample measurements from the SK screening effort, resulting in 15.9~kg$\cdot$yr of Gd$_2$(SO$_4$)$_3\cdot$8H$_2$O exposure, or 6.7~kg$\cdot$yr of natural Gd, on ULB HPGe detectors at Boulby. This represents the largest exposure of ULB gadolinium to date for conducting a search for rare Gd decays.

The BUGS facility houses several ULB HPGe detectors~\cite{bugslab}. These include "Merrybent" and "Belmont," both p-type coaxial detectors with masses of 2.0~kg and 3.2~kg, and relative efficiencies of 110\% and 160\%, respectively. Their background levels and resulting minimum detectable activities are reported in~\cite{bugsdetectors}.

\begin{figure}
    \centering
    \includegraphics[width=\linewidth]{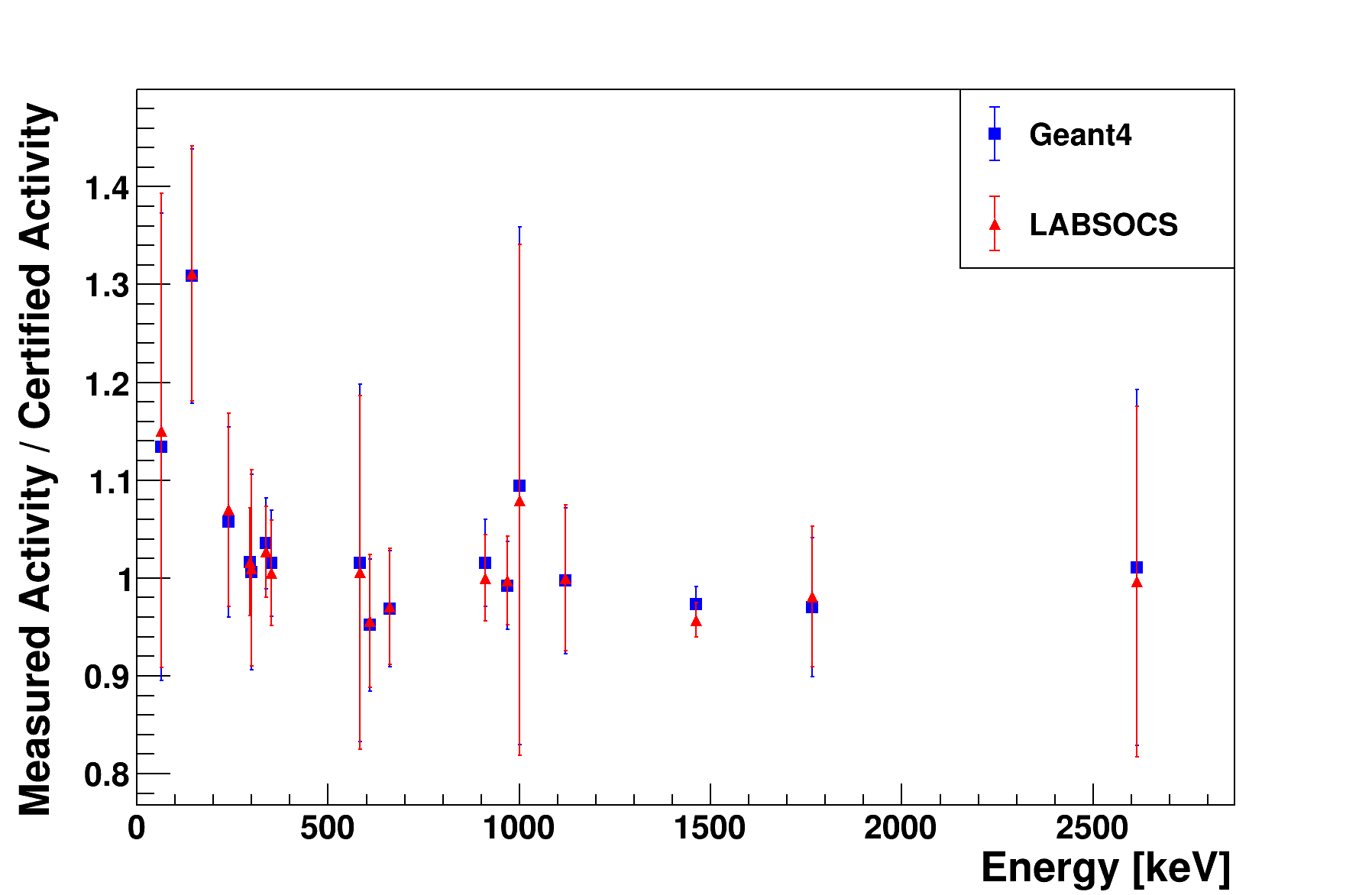}
    \caption{Results of a 3-day measurement of an IAEA-385 standard reference material on the Merrybent detector. The activities of various nuclides, derived using two independent calculations of detection efficiency and true coincidence summing factors, are compared with the certified values from the IAEA.}
    \label{fig:valMer}
\end{figure}

\begin{figure}
\centering
    \includegraphics[width=\linewidth]{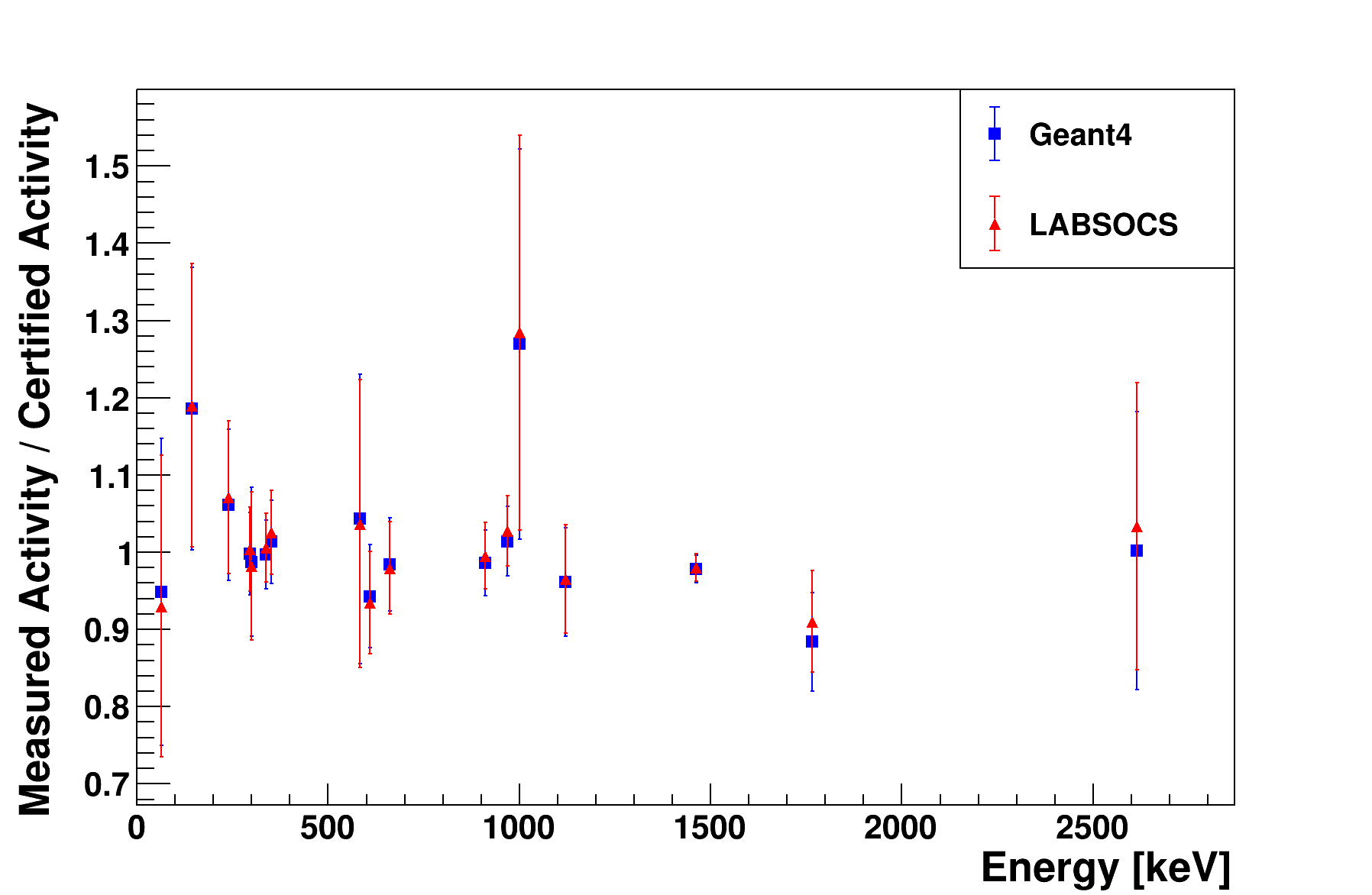}
    \caption{Results of a 3-day measurement of an IAEA-385 standard reference material on the Belmont detector. The activities of various nuclides, derived using two independent calculations of detection efficiency and true coincidence summing factors, are compared with the certified values from the IAEA.}
    \label{fig:valBel}
\end{figure}

The full energy peak efficiency and true coincidence summing factors for Merrybent and Belmont are calculated using a custom Geant4~\cite{geant4} simulation of the detectors, shielding, and sample geometry. This simulation, described in~\cite{boulby2018}, is validated against a LabSOCS~\cite{labsocs} efficiency simulation by measuring an IAEA-385 Irish Sea Sediment standard reference material on each detector. Both the Geant4 and LabSOCS calculated efficiencies and coincidence summing factors are applied to the IAEA-385 measurement and compared with the certified values. Good agreement is achieved across all analysed nuclides after accounting for coincidence summing and half-life corrections of the certified reference activities. Figures \ref{fig:valMer} and \ref{fig:valBel} demonstrate that the ratio of calculated reference source activities to the certified values is within $1\sigma$ of unity across all analysed nuclides. Consequently, the efficiencies calculated by the Geant4 simulation agree with those calculated using LabSOCS.

The Geant4 simulation package for Belmont and Merrybent was used to calculate the full energy peak efficiency for the expected gammas from the double beta decay of $^{160}$Gd to the $^{160}$Dy $0_2^+$, $0_1^+$, and $2_2^+$ levels, the alpha decay of $^{152}$Gd to the $^{148}$Sm $2_1^+$ level, and the alpha decay of $^{154}$Gd to the $^{150}$Sm $2_1^+$ level. The excited daughter states were generated, and the detected energy spectrum was analysed to determine the combined effect of detection efficiency and coincidence summing factors. The total efficiency factors for Belmont and Merrybent are reported in Table~\ref{tab:HLimits}.

\section{Cosmic Ray Activation of Natural Gd}

During its time on the Earth's surface and in transit from Japan to England, the Gd$_2$(SO$_4$)$_3\cdot$8H$_2$O material was exposed to high-energy cosmic rays. Potential long-lived radioisotope production was investigated using an ACTIVIA~\cite{activia} simulation of cosmogenic neutrons on natural Gd. While the activation of each measured batch varied, it generally consisted of several weeks of processing, storage, and transport at ground level, approximately one day in air transportation, and then several additional weeks on the surface before being transported underground. For modelling purposes, we simulated one month of activation at sea level, one day at 35,000~ft, and one additional month at sea level. Finally, a seven-day cooling-off period was simulated to represent the time spent underground at Boulby before screening commenced. ACTIVIA provides the corresponding neutron flux spectra as a function of atmospheric depth. After the activation and cooling-off periods, the remaining long-lived unstable products are isotopes of Eu, Gd, and Tb, as well as other isotopes produced through spallation and decay processes. Many of the most significant expected activation or spallation products with detectable gammas from the simulation are shown in Table~\ref{tab:activation}, with those having significant expected gammas indicated.
 
\begin{figure*}
  \centering
  \includegraphics[width=0.99\textwidth]{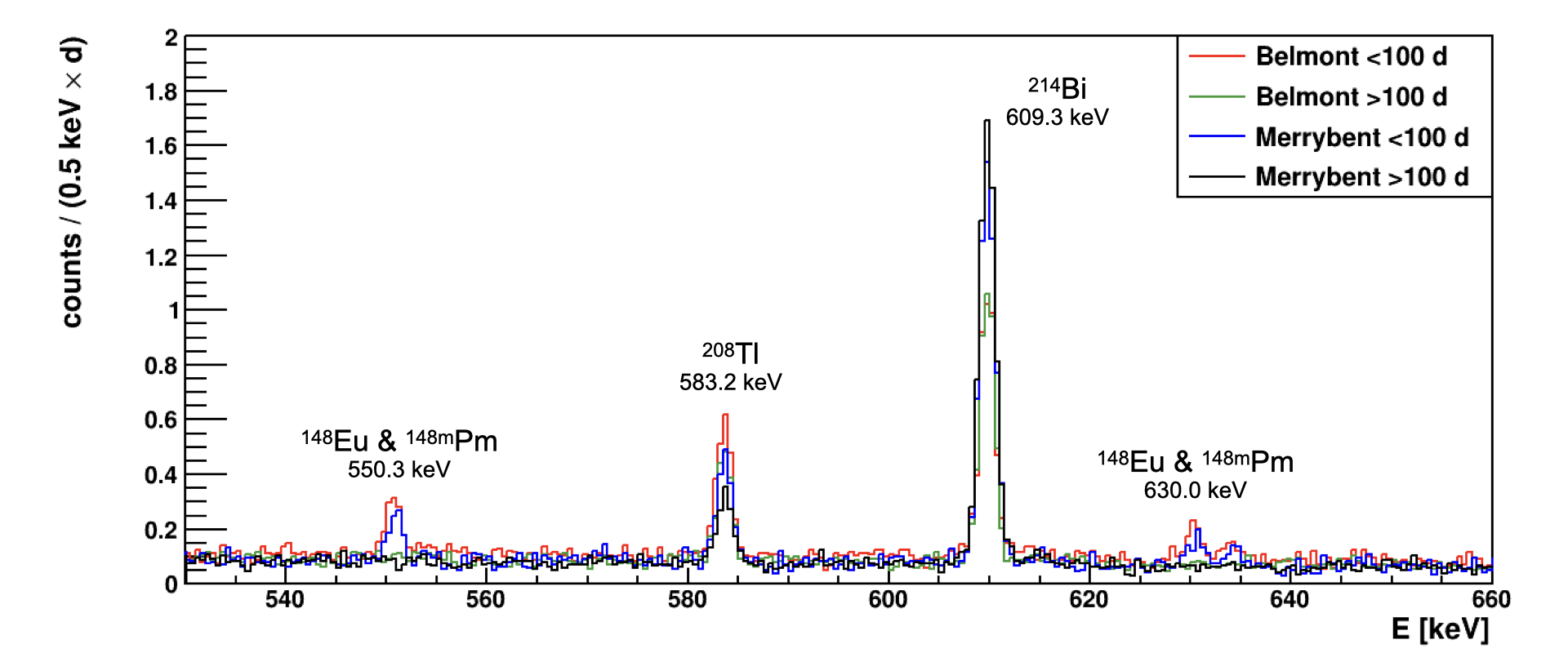}
  %\vspace{5cm}
\caption{\label{pic:DSSplitting}This figure illustrates the interference of cosmic activation peaks with the rare event search.  $^{148}$Eu ($T_{1/2} = 54.5$~d) and $^{148m}$Pm ($T_{1/2} = 41.29$~d) exhibit visible gamma lines at 550.3~keV and 630.0~keV, respectively. These interferences can be effectively eliminated by selecting data from samples that have spent more than 100 days underground.}
\end{figure*}

Activation products with gammas in the regions of interest for this rare Gd decay search are particularly problematic. Notably, $^{148}$Eu and $^{148m}$Pm have strong gammas (550.3~keV, I$_\gamma$\textgreater95\%, $T_{1/2} = 54.5$~d and $T_{1/2} = 41.29$~d, respectively~\cite{nds148}) that interferes with the $^{152}$Gd alpha decay search. The half-lives of these isotopes are similar in magnitude to the typical time between upper-atmosphere exposure to cosmic radiation and sample measurement on underground, ULB HPGe detectors. The residual activity of $^{148}$Eu following simulated activation indicates the potential for medium-lived gamma interference within these measured Gd$_2$(SO$_4$)$_3\cdot$8H$_2$O samples. This interference is most significant in samples screened soon after being brought underground. Both 550~keV and 630~keV gamma peaks are observed in the combined detector spectra (Figure~\ref{pic:DSSplitting}), with relative intensities consistent with literature values. A possible interference at 550~keV from $^{220}$Rn is dismissed as the peak area would imply a $^{232}$Th-chain activity which is two orders of magnitude greater than that observed from $^{208}$Tl. 

Other identified activation products in the datasets include $^{146}$Eu, $^{154}$Eu, $^{54}$Mn, $^{95}$Tc, and $^{143}$Pm (see Table~\ref{tab:activation}). Although sulfate activation was not simulated with ACTIVIA, the light spallation product $^{7}$Be was observed in the $<100$~d datasets. Possible interferences around 835~keV between $^{95}$Tc, $^{54}$Mn, and $^{228}$Ac are currently under investigation.

Due to the inconsistent cosmic ray irradiation and "cooling-down" periods of the Gd$_2$(SO$_4$)$_3\cdot$8H$_2$O samples before HPGe screening, the screening results were separated based on whether the screening commenced less than or greater than 100~days after being transported underground and shielded from cosmic rays. This cutoff was chosen to minimise the impact of interfering gammas while retaining a significant portion of the sample exposure for the $^{152}$Gd alpha decay search. Figure~\ref{pic:DSSplitting} shows the relevant region for $^{148}$Eu gammas for the Belmont \textless100~d, Belmont \textgreater100~d, Merrybent \textless100~d, and Merrybent \textgreater100~d datasets. The 100-days cutoff effectively eliminates the small but significant gamma peaks from both strong gamma emitters.

\begin{table*}
\centering
\begin{tabular}{@{}lllll@{}} \hline
Isotope & \begin{tabular}[t]{@{}l@{}}Half-Life \\ (days)\end{tabular} & \begin{tabular}[t]{@{}l@{}}Residual Activity\\ (mBq/kg)\end{tabular} & Significant Gammas & \begin{tabular}[t]{@{}l@{}}Observed Activity\\ (mBq/kg)\end{tabular}\\ \hline
%$^{147}$Eu & 24.6 & 31.58 & 197 keV (24\%) \\
%$^{151}$Gd & 120 & 31.48 & None \\
$^{156}$Eu & 15.2 & 0.33 & 811.8 keV (9.7\%) & $<0.16$ \\
$^{149}$Gd & 9.28 & 0.27 & 346.7 keV (23.7\%) & $<0.10$ \\
%$^{155}$Tb & 5.32 & 20.45 & None \\
$^{156}$Tb & 5.4 & 0.20 & 534.3 keV (67\%) & $<0.02$ \\
%$^{146}$Gd & 48.3 & 11.79 & (114.7 \& 115.5 keV), (88.6\%, sum) \\
$^{148}$Eu(*) & 54.5 & 0.12 & 550.3 keV (99\%), 630.0 keV (71.9\%) & $0.08 \pm 0.01$ \\
%$^{149}$Eu & 93.1 & 9.47 & None \\
$^{160}$Tb & 72.4 & 0.08 & 879.4 keV (30\%) & $<0.08$ \\
$^{145}$Eu & 5.93 & 0.08 & 893.7 keV (66\%) & $<0.02$ \\
%$^{153}$Gd & 242 & 5.70 & None \\
$^{143}$Pm & 265 & 0.06 & 741.9 keV (38.5\%) & $0.06 \pm 0.02$ \\
%$^{153}$Tb & 2.34 & 4.34 & None \\
%$^{139}$Ce & 137.6 & 3.64 & 165.9 keV (80\%) \\
$^{131}$Ba & 11.5 & 0.04 & 496.3 keV (48\%) & $<0.03$ \\
$^{146}$Eu & 4.51 & 0.03 & 747.2 keV (98\%), 633-634 keV (sum: 81.8\%) & $0.04 \pm 0.01$ \\
%$^{127}$Xe & 36.4 & 2.39 & 202.9 keV (69\%) \\
%$^{140}$Nd & 3.37 & 1.67 & None \\
%$^{125}$I & 60.2 & 1.38 & None \\
%$^{155}$Eu & 1737 & 1.33 & 86.5 keV (30.7\%) \\
%$^{121m}$Te & 154 & 0.98 & 212.2 keV (81.5\%) \\
%$^3$H & 4503.4 & 0.96 & None \\
%$^{145}$Sm & 340 & 0.95 & None \\
$^7$Be & 53.3 & 0.01 & 477.6 keV (10.4\%) & $0.39 \pm 0.09$ \\
$^{154}$Eu & 2993 & 0.009 & 1274.4 keV (34.8\%) & $0.09 \pm 0.03$ \\
$^{105}$Ag & 41.3 & 0.008 & 344.6 keV (42\%) & $<0.05$ \\
%$^{103}$Pd & 17 & 0.70 & None \\
%$^{134}$Ce & 3.16 & 0.64 & None \\
$^{144}$Pm & 363 & 0.007 & 618.0~keV (98\%), 696.5~keV (99.5\%) & $<0.03$\\
$^{113}$Sn & 115 & 0.006 & 391.7 keV (65\%) & $<0.02$ \\
%$^{118}$Te & 6 & 0.49 & None \\
$^{119}$Te & 4.68 & 0.005 & 1212.7 keV (66\%) & $<0.02$ \\
$^{152}$Eu & 4891 & 0.005 & 1408.0 keV (15\%), 964.1 keV (10.5\%) & $<0.19$\\
$^{85}$Sr & 64.8 & 0.005 & 514.0 keV (96\%) & $<0.02$ \\
%$^{71}$Ge & 11.2 & 0.39 & None \\
$^{88}$Zr & 83.4 & 0.004 & 392.9 keV (97.3\%) & $<0.03$ \\
$^{99}$Rh & 15 & 0.004 & 528.2 keV (37.9\%), 353.1 keV (34.5\%) & $<0.11$ \\
$^{95}$Tc & 61 & 0.004 & 582.1 keV (30\%), 835.1 keV (26.6\%) & $0.29 \pm 0.05$\\
%$^{73}$As & 80 & 0.31 & None \\
%$^{91}$Nb & 62 & 0.30 & None \\
$^{106}$Ag & 8.5 & 0.003 & 1045.8 keV (29.6\%) & $<0.08$\\
%$^{51}$Cr & 27.7 & 0.27 & 320.1 keV (9.9\%) \\
%$^{82}$Sr & 25 & 0.26 & None \\
%$^{32}$P & 14.3 & 0.25 & None \\
$^{148m}$Pm(*) & 41.3 & 0.003 & 550.3 keV (94.9\%), 630.0 keV (89\%) & $0.09 \pm 0.01$ \\ 
%$^{128}$Ba & 2.43 & 0.24 & 273.4 keV (14.5\%) \\
$^{54}$Mn & 312 & 0.001 & 834.8 keV (99.98\%) & $0.08 \pm 0.01$ \\ \hline
\end{tabular}\\
(*) The observed activity assumes all counts attributed to this nuclide.
\caption{A selection of long-lived radioisotopes produced by cosmic ray activation on natural Gd, as simulated by ACTIVIA. The residual specific activity for each nuclide is shown, considering one month of activation at sea level, one day at 35,000~ft, one additional month at sea level, and a final seven-day cooling-off period underground. Only products with long half-lives and significant gamma rays above 300~keV~\cite{NuclData} are included, as these are most likely to affect HPGe measurements of rare Gd decays. The final column shows the observed activity in the Belmont $<100$~d sample. Possible interferences, such as near 835~keV ($^{54}$Mn, $^{95}$Tc, $^{228}$Ac), have not been factored into the observed activities.}
\label{tab:activation}
\end{table*}

\section{Sample Characterization}

The concentration of chemical impurities in a subset of the Gd$_2$(SO$_4$)$_3\cdot$8H$_2$O samples was determined using a quadrupole inductively coupled plasma mass spectrometer equipped with collision cell (Agilent model 7850). The results are presented in Table~\ref{tab:GdICPMS}. A semi-quantitative analysis was performed, calibrating the instrument with a single standard solution containing 1~ppb of Li, Co, Y, Ce, and Tl. The uncertainty in the measured concentrations is approximately 25\%. The contamination of K was measured by high-resolution inductively coupled plasma mass spectrometry (HR-ICP-MS, Thermo Fisher Scientific ELEMENT2) in ``cool plasma'' mode in order to enhance the sensitivity. 

For each sample, 150~mg of Gd-containing material was placed in a plastic vial with 5~ml of ultrapure water and 0.5~ml of nitric acid. This mixture was then placed in an ultrasonic bath at $60^\circ$C until complete sample decomposition. The resulting solutions were diluted with ultrapure water to a total volume of 10~ml (a dilution factor of approximately 2000) in preparation for ICP-MS analysis.

\begin{table*}[htbp]
\centering
\begin{tabular}{c c c c c c c c c} 
 \hline
  Element & 190502 & 190804 & 220471 & 210601 & 190705 & 210811 & 210711 & 190904 \\ 
 \hline
 K & 380 & 295 & 200 & 635 & 440 & 425 & 300 & 210 \\
 Ba & 6 & 7 & 8 & 6 & 18 & 17 & 8 & 8 \\ 
  La & 147 & 60 & $<$ 10 & 2 & 58 & $<$ 10 & $<$ 20 & 35 \\
 Ce & 8 & 5 & $<$ 5 & $<$ 3 & 13 & $<$ 2 & $<$ 5 & 4 \\
 Nd & 40 & $<$ 10 & $<$ 20 & $<$ 20 & 120 & $<$ 50 & $<$ 50 & $<$ 30 \\ 
 Pr & 12 & 2 & 7 & 2 & 40 & 1 & 1 & 2 \\
 Sm & 185 & 230 & $<$ 20 & 6 & 215 & $<$20 & $<$ 50 & 220 \\
 Eu & 95 & 200 & 6 & 5 & 685 & 12 & 13 & 170 \\
 Tb & 135 & 140 & 70 & 137 & 96 & 130 & 110 & 97 \\
 Lu & $<$ 3000 & $<$ 3000 & $<$ 3000 & $<$ 3000 & $<$ 3000 & $<$ 3000 & $<$ 3000 & $<$ 3000 \\
 Th & $<$ 1 & $<$ 1 & $<$ 1 & $<$ 1 & $<$ 1 & $<$ 1 & $<$ 1 & $<$ 1 \\
 U & $<$ 1 & $<$ 1 & $<$ 1 & $<$ 1 & $<$ 1 & $<$ 1 & $<$ 1 & $<$ 1 \\
 \hline
\end{tabular}
\caption{\label{tab:GdICPMS}Concentrations, in units of parts-per-billion, of selected chemical impurities in Gd$_2$(SO$_4$)$_3\cdot$8H$_2$O samples, as determined by HR-ICP-MS analysis at LNGS.  The uncertainty in the measured concentrations is approximately 25\%. Limits are 68\% C.L. \ .}
\end{table*}

The results in Table~\ref{tab:GdICPMS} demonstrate the high chemical purity of the Gd-containing samples, confirming the effectiveness of the purification process employed during production. For example, the purification is highly effective at removing U and Th, with upper limits below 1~ppb established. The sensitivity of the ICP-MS measurements for these two elements could be further improved by employing a pre-concentration technique using anion-exchange resins~\cite{Ito2017,Nisi17}.

The high detection limit for Lu (3000~ppb) is caused by significant interference of signals from this element with the signal from the matrix (primarily GdOH$^+$ ions). Similarly, the signal from Tb is affected by interference from GdH$^+$ ions.

The radiopurity of all Gd$_2$(SO$_4$)$_3\cdot$8H$_2$O powder samples used in this analysis was previously reported by SK~\cite{gdproduction,secondgdloading}. For comparison with the ICP-MS measurements performed in this study, a subset of ULB HPGe-measured radiopurity values is reproduced in Table~\ref{tab:RadioGd}.

\begin{table*}[htbp]
    \centering
    \begin{tabular}{rrrrrrrrrr}
         \hline
         Chain	& Nuclide & \multicolumn{8}{c}{Specific activity, mBq/kg, 95\%~C.L. limits}\\
                \hline
            &       & 190502 & 190804 & 220471 & 210601 & 190705 & 210811 & 210711 & 190904 \\
            \hline
$^{238}$U	& $^{234}$Th & $< 4.5$ & $< 4.8$ & $< 6.8$ & $< 5.8$ & $9\pm5$ & $< 4.1$ & $< 8.2$ & $< 5.0$\\
        	& $^{214}$Pb & $< 0.20$ & $< 0.21$ & $0.86\pm0.23$ & $< 0.27$ & $< 0.30$ & $0.97\pm0.21$ & $< 0.23$ & $< 0.25$\\
%$^{238}$U   &$^{234}$Th	& $<0.2$ & Bq/kg\\
%            &$^{234m}$Pa& $<0.2$ & Bq/kg\\
%            &$^{226}$Ra	& $5 \pm 2$ & mBq/kg\\
$^{235}$U	& $^{235}$U	& $< 0.14$ & $< 0.27$ & $< 0.45$ & $< 0.10$ & $< 0.24$ & $< 0.18$ & $< 0.12$ & $< 0.26$\\
        	& $^{227}$Th & $< 0.62$ & $< 0.88$ & $< 0.75$ & $< 0.44$ & $1.4\pm0.5$ & $< 0.59$ & $< 0.58$ & $< 0.73$\\
%$^{235}$U	&$^{235}$U	& $<19$ & mBq/kg\\
%            &$^{231}$Pa	& $0.52 \pm 0.06$ &  Bq/kg\\
%            &$^{227}$Ac	& $18 \pm 4$ & mBq/kg\\
$^{232}$Th	& $^{228}$Ac & $< 0.41$ & $0.70\pm0.26$ & $0.85\pm0.34$ & $< 0.17$ & $0.30\pm0.18$ & $< 0.34$ & $< 0.40$ & $0.77\pm0.28$\\
        	& $^{208}$Tl & $< 0.37$ & $0.43\pm0.16$ & $0.96\pm0.28$ & $< 0.34$ & $0.53\pm0.16$ & $< 0.30$ & $< 0.22$ & $0.81\pm0.22$\\
%$^{232}$Th	&$^{228}$Ra	& $<6$ & mBq/kg\\
%            &$^{228}$Th	& $<6$ & mBq/kg\\ %m
            
            \\
            & $^{176}$Lu & $0.34\pm0.08$ & $5.1\pm0.3$ & $0.12\pm0.06$ & $0.69\pm0.09$ & $1.8\pm0.2$ & $0.29\pm0.06$ & $0.37\pm0.06$ & $7.1\pm0.4$\\
            & $^{137}$Cs & $< 0.04$ & $< 0.05$ & $< 0.05$ & $< 0.02$ & $< 0.08$ & $< 0.08$ & $< 0.04$ & $< 0.09$\\
            & $^{138}$La & $0.12\pm0.07$ & $< 0.13$ & $< 0.07$ & $< 0.09$ & $< 0.14$ & $< 0.14$ & $< 0.10$ & $< 0.13$\\
            %& $^{60}$Co & $< 0.13$ & $< 0.13$ & $< 0.12$ & $< 0.10$ & $< 0.09$ & $< 0.09$ & $< 0.08$ & $< 0.14$\\
            & $^{40}$K	& $< 2.7$ & $< 3.5$ & $< 6.3$ & $< 0.98$ & $< 1.7$ & $< 1.8$ & $< 1.4$ & $< 4.0$\\
            \hline
    \end{tabular}
    \caption{\label{tab:RadioGd}Specific activities of naturally occurring and cosmogenically activated radionuclides in the selected Gd$_2$(SO$_4$)$_3\cdot$8H$_2$O powder samples, as measured by gamma-ray spectrometry.}
\end{table*}

The upper detection limit for \nuc{K}{40}, less than (1--6)~mBq/kg, corresponds to a concentration of natural potassium of less than 32--192~ppb. These are in slight disagreement with the concentrations of natural potassium reported by ICP-MS. The BUGS laboratory is located within the potash layer of the Boulby Mine, so the potassium background is greater and sensitivities worse compared to other underground sites around the world.

Similarly, the observed specific activity of \nuc{La}{138} in sample 190502 (0.12~mBq/kg) corresponds well with the highest concentration of La (approximately 150~ppb) among all samples measured using ICP-MS.

\nuc{Lu}{176} decays through beta decay with gamma emission. The detected activity of this radioisotope in all analysed samples, at the level of (0.12--7.11)~mBq/kg, corresponds to a natural lutetium concentration of 2.3--137.0~ppb. This result agrees well with the established detection limit for Lu in the ICP-MS measurements (\textless3000~ppb). However, due to the strong interference effects with chemical elements of the complex compound observed with ICP-MS, gamma-ray spectrometry provides a more precise evaluation of the Lu concentration in Gd$_2$(SO$_4$)$_3\cdot$8H$_2$O samples.

It is important to note that chemical reactions and transformations typically disrupt the secular equilibrium in natural radioactive decay chains. This disruption is primarily driven by the chemical properties of the daughter nuclides in the U/Th chains. Furthermore, the production technology for the Gd$_2$(SO$_4$)$_3\cdot$8H$_2$O material used in SK was modified at least twice, leading to different disruptions of the secular equilibrium in each production run\footnote{Samples numbered 19xxxx were from the first production run. Samples numbered 21xxxx or 22xxxx were from the second production run, where radium was reduced further compared with the earlier batches.}. This variation can be observed in the specific activities of the radionuclides from the natural U/Th decay chains.

The isotopic composition of gadolinium was determined through complementary measurements of the eight Gd$_2$(SO$_4$)$_3\cdot$8H$_2$O samples selected for comparative chemical and radiopurity studies. HR-ICP-MS analysis was conducted at LNGS, and the results are presented in Table~\ref{tab:GdIsotopic}. The measured isotopic composition of gadolinium in all eight samples is consistent with natural abundance, within the quoted uncertainties. Therefore, tabulated values for the natural isotopic composition of gadolinium were used in subsequent analyses.

\begin{table*}[htbp]
\centering
\begin{tabular}{c c c c c c c c} 
 \hline
  Sample & \nuc{Gd}{152} & \nuc{Gd}{154} & \nuc{Gd}{155} & \nuc{Gd}{156} & \nuc{Gd}{157} & \nuc{Gd}{158} & \nuc{Gd}{160} \\ 
 \hline
 190502 & $0.18\pm0.01$ & $2.20\pm0.01$ & $14.80\pm0.10$ & $20.49\pm0.18$ & $15.63\pm0.08$ & $24.77\pm0.02$ & $21.92\pm0.12$ \\
 190804 & $0.18\pm0.01$ & $2.18\pm0.02$ & $14.80\pm0.05$ & $20.37\pm0.10$ & $15.50\pm0.06$ & $24.96\pm0.05$ & $21.97\pm0.10$ \\ 
 220471 & $0.20\pm0.01$ & $2.18\pm0.01$ & $14.90\pm0.11$ & $20.43\pm0.25$ & $15.64\pm0.13$ & $24.66\pm0.17$ & $21.95\pm0.12$ \\
 210601 & $0.18\pm0.01$ & $2.18\pm0.01$ & $14.70\pm0.03$ & $20.53\pm0.08$ & $15.79\pm0.05$ & $24.74\pm0.02$ & $21.84\pm0.11$ \\
 190705 & $0.20\pm0.01$ & $2.17\pm0.01$ & $14.80\pm0.25$ & $20.50\pm0.03$ & $15.66\pm0.06$ & $24.82\pm0.33$ & $21.86\pm0.15$ \\
 210811 & $0.18\pm0.01$ & $2.20\pm0.02$ & $14.70\pm0.14$ & $20.52\pm0.26$ & $15.72\pm0.11$ & $24.89\pm0.23$ & $21.82\pm0.02$ \\
 210711 & $0.20\pm0.01$ & $2.19\pm0.01$ & $14.70\pm0.04$ & $20.51\pm0.05$ & $15.63\pm0.05$ & $25.03\pm0.13$ & $21.74\pm0.07$ \\ 
 190904 & $0.20\pm0.01$ & $2.18\pm0.02$ & $14.70\pm0.10$ & $20.43\pm0.21$ & $15.64\pm0.16$ & $25.06\pm0.19$ & $21.78\pm0.21$ \\
 \hline
 Tabulated value & $0.20\pm0.01$ & $2.18\pm0.02$ & $14.80\pm0.09$ & $20.47\pm0.03$ & $15.65\pm0.04$ & $24.84\pm0.08$ & $21.86\pm0.03$ \\
 \hline
\end{tabular}
\caption{\label{tab:GdIsotopic}The isotopic abundances of gadolinium in the eight Gd$_2$(SO$_4$)$_3\cdot$8H$_2$O samples, as determined by HR-ICP-MS analysis.}
\end{table*}

\section{Datasets}

Gd$_2$(SO$_4$)$_3\cdot$8H$_2$O samples were measured at BUGS from September 2019 to February 2024 (Table~\ref{tab:datasets} in appendix). Each sample was screened until achieving a sensitivity of \textless0.5~mBq/kg for the 609.3~keV gamma ray from $^{214}$Bi, at a 95\% confidence level. In some instances, such as during the first UK COVID-19 lockdown from March to May 2020, sample measurements continued beyond the time required to reach this sensitivity due to restricted underground access. The inherent radioactivity of each batch is reported in~\cite{gdproduction} and \cite{secondgdloading}, relative to the SK requirements. This Gd$_2$(SO$_4$)$_3\cdot$8H$_2$O material was among the most radiopure ever screened at Boulby, necessitating additional corrections to the spectral analysis to account for the shielding of background sources by the large, dense samples~\cite{bkgshield}.

To account for drift in the absolute ADC calibration, fluctuating backgrounds, and external factors affecting detector resolution, each measured spectrum was individually calibrated. This calibration procedure involved identifying significant peaks attributable to common naturally occurring radioactive material (NORM) gamma full energy peaks. Depending on the Gd batch, these peaks could include 201.83~keV and 306.82~keV from $^{176}$Lu, 295.22~keV and 351.93~keV from $^{214}$Pb, 609.31~keV, 1120.29~keV, 1764.49~keV, and 2204.21~keV from $^{214}$Bi, 911.20~keV and 968.96~keV from $^{228}$Ac, 583.19~keV and 2614.51~keV from $^{208}$Tl, and 1173.21~keV and 1332.49~keV from $^{60}$Co. A linear ADC calibration was evaluated for each spectrum based on the locations of these peaks and the literature gamma energy values. The Gaussian width parameter of each fitted peak was also used to estimate the detector energy resolution, $\sigma(E)$, with the functional form: $\sigma(E)=a+b\sqrt{E}$, where $a$ and $b$ are free parameters. The resolution for each detector remained stable throughout the screening programme. The corresponding values for the gamma energies of interest for this analysis are shown in Table~\ref{tab:HLimits}.

Detector stability was monitored throughout the Gd measurement period using several methods. The inherent background spectrum, measured multiple times during the screening programme, was checked for consistency. Periods exhibiting higher-than-normal background levels were investigated and generally attributed to external factors, such as N$_2$ purge failures. Additionally, periodic measurements of a check source containing 37~kBq of both $^{155}$Eu and $^{22}$Na were conducted to monitor the full energy peak width of the characteristic gamma rays and the dead layer thickness. The latter was determined by examining the ratio of the count rate in the 86.5~keV and 105.3~keV peak areas. Datasets exhibiting fluctuations in detector background or full energy peak resolution indicative of suboptimal detector conditions were excluded.

\begin{figure*}
  \centering
  \includegraphics[width=0.99\textwidth]{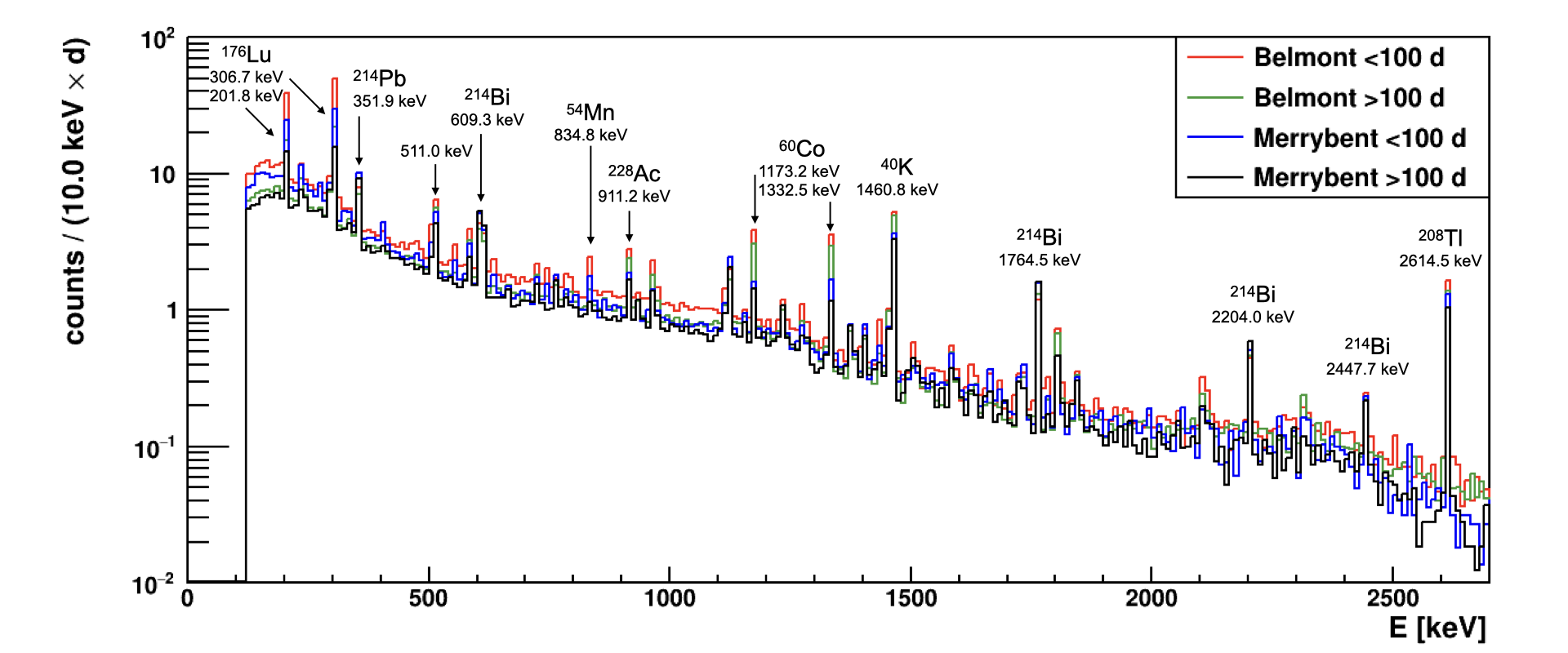}
  %\vspace{5cm}
\caption{\label{pic:Spectrum}This figure presents the spectra of all four datasets used in this analysis, categorized by detector (Belmont and Merrybent) and the time elapsed between the samples being brought underground and the commencement of measurements (\textless{100}~days and \textgreater{100}~days).}
\end{figure*}

The calibrated spectra from each detector were rebinned and combined according to the \textless100~day and \textgreater100~day classifications. Figure \ref{pic:Spectrum} displays the full spectra in 10 keV bins for all four combined datasets. Statistical analysis of the rare event search employed 0.5~keV binned spectra.

%%%%%%%%%%%%%%%%%%%%%%%%%%%%%%%%%%%%%%%%%%%%%%%%%%%%%%%%%%%%%
%%%%%%%%%%%%%%%%%%%%%%%%%%%%%%%%%%%%%%%%%%%%%%%%%%%%%%%%%%%%%
%%%%%%%%%%%%%%%%%%%%%%%%%%%%%%%%%%%%%%%%%%%%%%%%%%%%%%%%%%%%%
\section{Analysis}

This analysis builds upon the work presented in~\cite{Laubenstein23}, employing the same peak search procedure based on Bayesian statistics (see Eqs.\ (5), (6), and (7) from that manuscript) and incorporating the results from the same study as prior input. For this study, as in the previous one, an independent analysis is performed for each decay mode. All four datasets are combined using fits with a single inverse half-life parameter $(T_{1/2})^{-1}$ but independent nuisance parameters. To avoid interference from cosmogenically produced isotopes ($^{148}$Eu and $^{148m}$Pm), only the two datasets with \textgreater100~days underground are used for the alpha decay of \nuc{Gd}{152}. The marginalized posteriors of $(T_{1/2})^{-1}$ from \cite{Laubenstein23} serve as prior functions for $(T_{1/2})^{-1}$ in this analysis.

The live times, $T$, of the four datasets were as follows: (i) 0.81919~yr for Belmont \textless{100}~d; (ii) 1.03781~yr for Belmont \textgreater{100}~d; (iii) 0.60820~yr for Merrybent \textless{100}~d; and (iv) 0.88705~yr for Merrybent \textgreater{100}~d. The mass of the gadolinium sample was 4.78~kg for all samples and datasets, corresponding to $1.54\times10^{22}$, $1.68\times10^{23}$, $3.37\times10^{24}$ atoms of \nuc{Gd}{152}, \nuc{Gd}{154}, and \nuc{Gd}{160}, respectively. The uncertainties associated with sample mass and isotopic abundance are subdominant and correlated to the uncertainty of the detection efficiency (10\%); therefore, they are neglected. The detection efficiencies and energy resolutions, which differ by detector but share the same prior input for the \textless{100}~d and \textgreater{100}~d division, are provided in Table \ref{tab:HLimits}. It is important to note that each dataset has its own fit parameters, allowing for systematic variations between datasets even with identical input values.

\begin{figure*}
  \centering
  \includegraphics[width=0.99\textwidth]{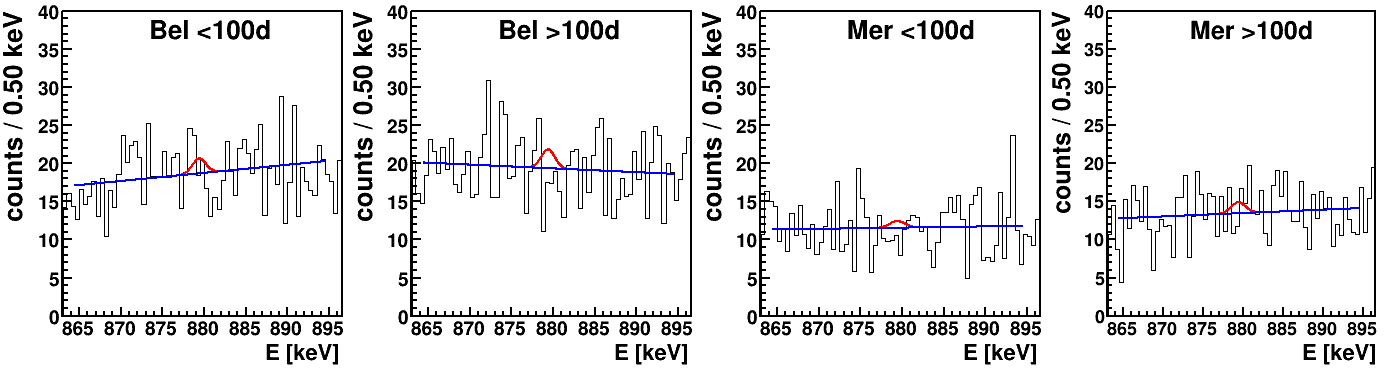}
\caption{Fit of the 879.4~keV \gline\ from \nuc{Gd}{160} $0\nu/2\nu\beta\beta$ decay into the excited $2^+_2$ state (966.2~keV) of \nuc{Dy}{160}. }
  \label{pic:pdf_ROI_Composition_DBD_Gd160_bb_2p2_GdSK_4DS_v5}
\end{figure*}

The spectral fits for all decay modes are shown in Figures \ref{pic:pdf_ROI_Composition_DBD_Gd160_bb_2p2_GdSK_4DS_v5}–\ref{pic:pdf_ROI_Composition_Alpha_Gd154_aES_GdSK_4DS_v5} in the Appendix. The best-fit functions are shown in blue. All investigated rare decay modes are consistent with the assumption of no detected signal. The fit functions with the signal process set to the 90\% credibility limit are shown in red.

The 90\% quantiles of the marginalized posteriors of $T_{1/2}^{-1}$ yield the limits. These distributions are shown in red in Figure \ref{pic:posteriors} in the Appendix. The blue curves represent the priors from~\cite{Laubenstein23}, and the shaded areas indicate the 90\% probability regions of the posterior.

The lower half life limits are presented in the last columns of Table \ref{tab:HLimits}, showing the previous limits from~\cite{Laubenstein23}, the limits obtained using only data from this work (flat prior in $T_{1/2}^{-1}$), and the combined data as described above.

The data combination results in both higher and lower limits compared to using the new data alone but does not significantly influence the overall results. This outcome is attributed to the significantly greater amount of data and sensitivity in the present work, resulting in a relatively flat and uninformative prior. Random over- or under-fluctuations in the previous data only slightly inform the Bayesian parameter interpretation of the new data. Nevertheless, using the combined results is recommended, as they utilise more experimental data and avoid bias from selecting the best limit. Therefore, the reported posteriors in Figure \ref{pic:posteriors} can serve as prior input for future searches.

\begin{table*}[htbp]
\centering
\begin{tabular}{llllllll}
\hline
Nuclide (decay) & Daughter (level)  &    \glines  &  Det.\ eff.\ $\epsilon$   & $\sigma_{\rm res}$ &    T$_{1/2}$ (90\% C.I.)  &  T$_{1/2}$ (90\% C.I.) & T$_{1/2}$ (90\% C.I.) \\
                         &  ($J^\pi $ keV)  &     [keV]                    &  [\%]               &  [keV]                      &    [yr]  old \cite{Laubenstein23} &    [yr] new  &    [yr] comb. \\
\hline

%\nuc{Gd}{152} (\bbe{0}) & \nuc{Sm}{152} ($0^+_{\rm g.s.})$   &   39.5-46.5    &  \baseTsolo{-6} - \baseTsolo{-4}   &   0.60  &   \baseT{>4.2}{12}               \\

%\nuc{Gd}{160} $(0\nu/2\nu\beta\beta)$ & \nuc{Dy}{160} ($2^+_1 86.8)$   &   86.8    & 0.088 &   0.66  &   \baseT{>1.8}{18}               \\
\nuc{Gd}{160} $(0\nu/2\nu\beta\beta)$  & \nuc{Dy}{160} ($2^+_2 966.2)$   &   879.4    & 1.23,  0.84  &   0.84, 0.95  &    \baseT{>9.7}{19}  (*)  &  \baseT{>1.4}{21}  &  \baseT{>1.5}{21}           \\
\nuc{Gd}{160} $(0\nu/2\nu\beta\beta)$ & \nuc{Dy}{160} ($0^+_1 1279.9)$   &   1193.2    & 2.06, 1.37    &   0.90, 1.03 &   \baseT{>8.2}{19}    &   \baseT{>2.7}{21}  &  \baseT{>2.7}{21}      \\
\nuc{Gd}{160} $(0\nu/2\nu\beta\beta)$  & \nuc{Dy}{160} ($0^+_2 1456.8)$   &   1369.9    & 1.89, 1.24   &   0.94, 1.08  &   \baseT{>5.0}{19}   &   \baseT{>4.2}{21}   &  \baseT{>4.3}{21}       \\

\hline
\nuc{Gd}{152} ($\alpha$) & \nuc{Sm}{148} ($2^+_1 550.3)$   &   550.3    & 2.56, 1.82   &   0.70, 0.78  &   \baseT{>3.4}{17}   &   \baseT{>9.0}{18}    &  \baseT{>8.4}{18}           \\
\nuc{Gd}{154} ($\alpha$) & \nuc{Sm}{150} ($2^+_1 333.9)$   &   333.9    & 2.64, 1.93   &   0.60, 0.67  &   \baseT{>9.6}{18}   &   \baseT{>9.7}{19}   &  \baseT{>1.0}{20}           \\

\hline
\end{tabular}
\begin{flushleft}(*) the old half-life limit is based on a combined fit of two \glines\end{flushleft}
%\medskip
\caption{\label{tab:HLimits}Lower half-life limits for the investigated decay modes in gadolinium isotopes. Columns 3-5 display the gamma lines used in the fit, along with their corresponding detection efficiencies and energy resolutions. The latter two columns are separated for the Belmont and Merrybent detectors.}
\end{table*}

%%%%%%%%%%%%%%%%%%%%%%%%%%%%%%%%%%%%%%%%%%%%%%%%%%%%%%%%%%%%%
%%%%%%%%%%%%%%%%%%%%%%%%%%%%%%%%%%%%%%%%%%%%%%%%%%%%%%%%%%%%%
%%%%%%%%%%%%%%%%%%%%%%%%%%%%%%%%%%%%%%%%%%%%%%%%%%%%%%%%%%%%%
\section{Discussion}

It is noteworthy that the sensitivity of modern experimental techniques, such as the ULB HPGe gamma spectrometers employed in this study, allows for the investigation of rare nuclear processes even within routine materials screening campaigns. This study demonstrates the feasibility of achieving experimental sensitivities of 10$^{19}-10^{21}$~years for various rare nuclear processes in natural Gd isotopes. This was accomplished using a set of large-mass Gd-containing samples with a total exposure of 6.7~kg$\cdot$yr of natural Gd, originally intended for radiopurity assays.

None of the investigated decay modes yielded an observable signal. 90\% credibility limits were established using a Bayesian analysis that accounted for dominant systematic uncertainties. The recently established experimental limits from \cite{Laubenstein23} were improved by approximately two orders of magnitude. However, these experimental limits for the alpha decays of \nuc{Gd}{152} and \nuc{Gd}{154} remain far from theoretical expectations ($10^{25}$ yr and $10^{80}$~yr, respectively). Calculations for the expected half-lives of various double beta decay modes of \nuc{Gd}{160} are still pending.

Further improvements in experimental searches for rare decays of \nuc{Gd}{152}, \nuc{Gd}{154}, and \nuc{Gd}{160} are challenging within the "source $\neq$ detector" approach. This limitation arises from two main disadvantages of this technique: (i) the decrease in detection efficiency with decreasing gamma energy due to absorption in surrounding materials (e.g., sample container, end cap, and Ge crystal holder); and (ii) the limited benefit of increasing sample mass due to increased self-absorption.

Therefore, innovative experimental approaches that significantly enhance detection efficiency, such as those described in~\cite{Nagorny21}, or increase the number of isotopes of interest through enrichment~\cite{Beeman2015} are necessary. The latter option appears more promising, as the potential sensitivity enhancement would be the product of the enrichment factor and the improvement in detection efficiency achieved through reduced sample dimensions and self-absorption. This enhancement could range from a factor of 5 to several orders of magnitude.

A more substantial enhancement in detection efficiency for the processes of interest could be achieved by implementing the "source = detector" approach. In this approach, the decaying Gd nuclei are embedded within the detector's sensitive volume, which could be a liquid scintillator, a crystal scintillator, or a bolometer/scintillating bolometer. This configuration enables the detection of not only de-excitation gamma quanta but also short-range particles (alpha and beta particles) in the final reaction channel. This capability allows for further background rejection through coincidence measurements of different reaction products.

Furthermore, effective particle identification based on pulse-shape discrimination, variations in emitted scintillation light, or variations in the ratio of emitted light to phonon signal would enable the detection of alpha decay to ground states. This capability would also enhance the experimental sensitivity to decay modes that occur through transitions to excited states of daughter nuclei. Finally, this experimental approach accommodates the use of a much larger sample mass.

For example, the PIKACHU (Pure Inorganic scintillator experiment in KAmioka for CHallenging Underground sciences) project was recently initiated to fabricate high-purity Ce-doped Gd$_{3}$Ga$_{3}$Al$_{2}$O$_{12}$ (GAGG - Gadolinium Aluminium Gallium Garnet) single crystals for studying all modes of \nuc{Gd}{160} double beta decay~\cite{omori2024first}. This experiment benefits from a combination of technological expertise and successful multi-ton Gd-salt purification for the SK experiment~\cite{gdproduction} and in-house growth of GAGG single crystals with strict control at each production stage. Moreover, crystal growth leverages the well-developed technology of GAGG single crystal production (see, for instance \cite{Kamada12} and \cite{Kamada16}), driven by commercial market demands. Consequently, high-quality, large-volume scintillating crystals with masses of a few kilograms are commercially available.

Initially, PIKACHU will employ two GAGG scintillating crystals, 6.5~cm in diameter and 14.5~cm in length, each containing 710~g of \nuc{Gd}{160}. Due to the excellent light yield (approximately 50,000~photons/MeV) and pulse-shape discrimination capability of these crystals, a sensitivity level of $10^{22}$ years is anticipated for the double beta decay modes of \nuc{Gd}{160}. Subsequently, twenty large-volume, high-radiopurity GAGG crystals will be used to further enhance the experimental sensitivity. Therefore, the most stringent constraints on not only \nuc{Gd}{160} double beta decay modes but also rare alpha decays of \nuc{Gd}{152} and \nuc{Gd}{154} will likely emerge from this experiment in the near future.

\section{Conclusion}

This study investigated rare nuclear decay modes in natural gadolinium isotopes using high-purity samples originally intended for the Super Kamiokande neutrino experiment.  Despite achieving a sensitivity of $10^{19}-10^{21}$ years, no evidence for alpha or double beta decay to excited states of daughter nuclides  was observed. The resulting limits improve upon previous measurements by two orders of magnitude. Future searches employing enriched isotopes or the "source = detector" approach are necessary to reach the theoretical predictions for these decays.  Projects such as PIKACHU, which leverage high radiopurity GAGG scintillating crystals, hold promise for achieving the required sensitivities and providing crucial insights into rare nuclear processes that may occur in Gd isotopes and neutrino properties.

%%%%%%%%%%%%%%%%%%%%%%%%%%%%%%%%%%%%%%%%%%%%%%%%%%%%%%%%%%%%%
%%%%%%%%%%%%%%%%%%%%%%%%%%%%%%%%%%%%%%%%%%%%%%%%%%%%%%%%%%%%%
%%%%%%%%%%%%%%%%%%%%%%%%%%%%%%%%%%%%%%%%%%%%%%%%%%%%%%%%%%%%%
\begin{acknowledgements}

The authors express their sincere gratitude to the Super-Kamiokande Collaboration for providing the Gd samples used in this study. This work was supported by the U.K. Science and Technology Facilities Council [grant numbers ST/T00200X/1, ST/V002821/1, ST/V006185/1, ST/X002438/1]. We would also like to thank Ruben Saakyan who suggested looking for rare decays in gadolinium screening data and connected our small collaboration. 

\end{acknowledgements}

\bibliographystyle{spphys}       % APS-like style for physics


%%%%%%%%%%%%%%%%%%%%%%%%%%%%%%%%%%%%%%%%%%%%%%%%%%%%%%%%%%%%%
%%%%%%%%%%%%%%%%%%%%%%%%%%%%%%%%%%%%%%%%%%%%%%%%%%%%%%%%%%%%%
%%%%%%%%%%%%%%%%%%%%%%%%%%%%%%%%%%%%%%%%%%%%%%%%%%%%%%%%%%%%%
%\newpage
\section{Appendix}

%\begin{comment}
% \usepackage{booktabs}
\begin{table*}
\centering
\begin{tabular}{@{}cccccc@{}}
\hline
\textbf{Sample} & \textbf{Detector} &  \textbf{\begin{tabular}[c]{@{}c@{}}Measurement\\Time \\ (days)\end{tabular}} & \textbf{\begin{tabular}[c]{@{}c@{}}Underground\\ Time\\ (days)\end{tabular}} & \textbf{Start Date} \\ \hline
190502 & Belmont &  35.2 & \textless{}~100 & 2019/09/24 \\
190604 & Belmont &  30.3 & \textgreater{}~100 & 2020/08/07 \\
190704 & Belmont &  37.3 & \textless{}~100 & 2019/11/29 \\
190706 & Belmont &  27.4 & \textless{}~100 & 2020/01/09 \\
190804 & Belmont &  31.0 & \textless{}~100 & 2020/02/14 \\
190902 & Belmont &  98.4 & \textless{}~100 & 2020/03/18 \\
%190902\_E & Belmont  & 52.9 & \textless{}100d & 2020/03/18 \\
%190902\_L & Belmont  & 45.5 & \textgreater{}100d & 2020/05/11 \\
190902\_2 & Belmont  & 7.4 & \textgreater{}~100 & 2021/06/29 \\
190902\_3 & Belmont  & 18.3 & \textgreater{}~100 & 2021/08/13 \\
190904 & Belmont & 28.4 & \textgreater{}~100 & 2020/07/09 \\
210601 (A) & Belmont  & 54.1 & \textgreater{}~100 & 2022/02/18 \\
210711 (C) & Belmont  & 31.7 & \textless{}~100 & 2021/12/16 \\
210712 (D) & Belmont  & 37.7 & \textless{}~100 & 2022/12/20 \\
210713\_2 (E) & Belmont  & 19.5 & \textgreater{}~100 & 2023/02/27 \\
210811 (F) & Belmont &  25.2 & \textgreater{}~100 & 2022/04/14 \\
211106\_2 (K) & Belmont  & 37.1 & \textgreater{}~100 & 2023/04/04 \\
220241 (M) & Belmont &  40.2 & \textgreater{}~100 & 2023/06/09 \\
220242 (N) & Belmont &  25.7 & \textgreater{}~100 & 2023/07/20 \\
220251 (O) & Belmont &  36.9 & \textgreater{}~100 & 2023/09/15 \\
220352 (Q) & Belmont &  20.4 & \textgreater{}~100 & 2023/11/01 \\
220361 (S) & Belmont &  15.2 & \textgreater{}~100 & 2023/11/22 \\
220471 (U) & Belmont &  20.4 & \textgreater{}~100 & 2024/01/25 \\
190501 & Merrybent &  12.7 & \textless{}~100 & 2019/09/10 \\
190606 & Merrybent &  28.5 & \textless{}~100 & 2019/10/24 \\
190705 & Merrybent &  44.1 & \textless{}~100 & 2019/12/19 \\
190705\_2 & Merrybent  & 24.9 & \textgreater{}~100 & 2021/07/01 \\
190802 & Merrybent &  56.9 & \textgreater{}~100 & 2020/07/09 \\
190806 & Merrybent &  45.1 & \textless{}~100 & 2020/03/11 \\
210601 (A) & Merrybent  & 9.2 & \textless{}~100 & 2021/09/02 \\
210711 (C) & Merrybent  & 93.0 & \textgreater{}~100 & 2022/11/17 \\
210811 (F) & Merrybent  & 61.3 & \textgreater{}~100 & 2023/06/09 \\
210821 (G) & Merrybent  & 82.7 & \textless{}~100 & 2021/12/16 \\
%210821\_E (G) & Merrybent  & 49.2 & \textless{}100d & 2021/12/16 \\
%210821\_L (G) & Merrybent  & 33.5 & \textgreater{}100d & 2022/02/09 \\
210822 (I) & Merrybent &  20.2 & \textgreater{}~100 & 2023/11/01 \\
211201\_2 (L) & Merrybent & 20.3 & \textgreater{}~100 & 2023/11/22 \\
220482 (W) & Merrybent &  24.4 & \textgreater{}~100 & 2024/01/04 \\
220582 (Y) & Merrybent &  12.3 & \textgreater{}~100 & 2024/01/31 \\
220691 (Z) & Merrybent &  10.5 & \textgreater{}~100 & 2024/02/13 \\ \hline
\end{tabular}
\caption{Constituent datasets used in the current analysis.}
\label{tab:datasets}
\end{table*}
%\end{comment}

\begin{figure*}
  \centering
  \includegraphics[width=0.99\textwidth]{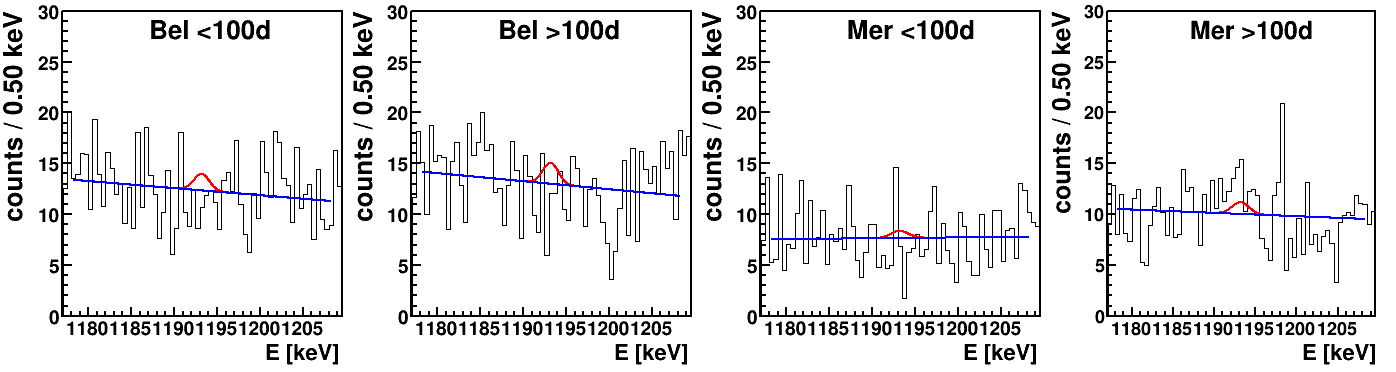}
\caption{Fit of the 1193.2~keV \gline\ from \nuc{Gd}{160} $0\nu/2\nu\beta\beta$ decay into the excited $0^+_1$ state (1279.9~keV) of \nuc{Dy}{160}. }
  \label{pic:pdf_ROI_Composition_DBD_Gd160_bb_0p1_GdSK_4DS_v5}
\end{figure*}

\begin{figure*}
  \centering
  \includegraphics[width=0.99\textwidth]{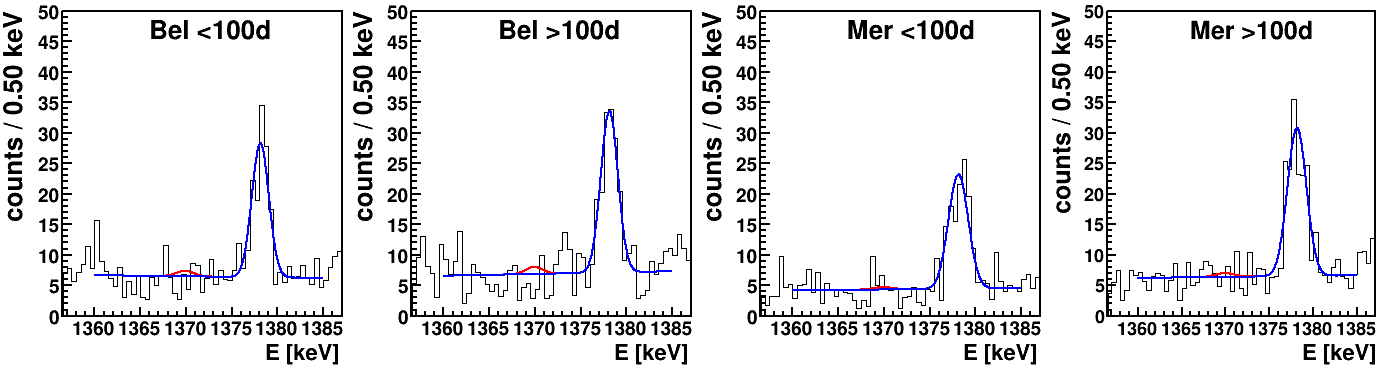}
\caption{Fit of the 1369.9~keV \gline\ from \nuc{Gd}{160} $0\nu/2\nu\beta\beta$ decay into the excited $0^+_2$ state (1456.8~keV) of \nuc{Dy}{160}.}
  \label{pic:pdf_ROI_Composition_DBD_Gd160_bb_0p2_GdSK_4DS_v5}
\end{figure*}

\begin{figure*}
  \centering
  \includegraphics[width=0.99\textwidth]{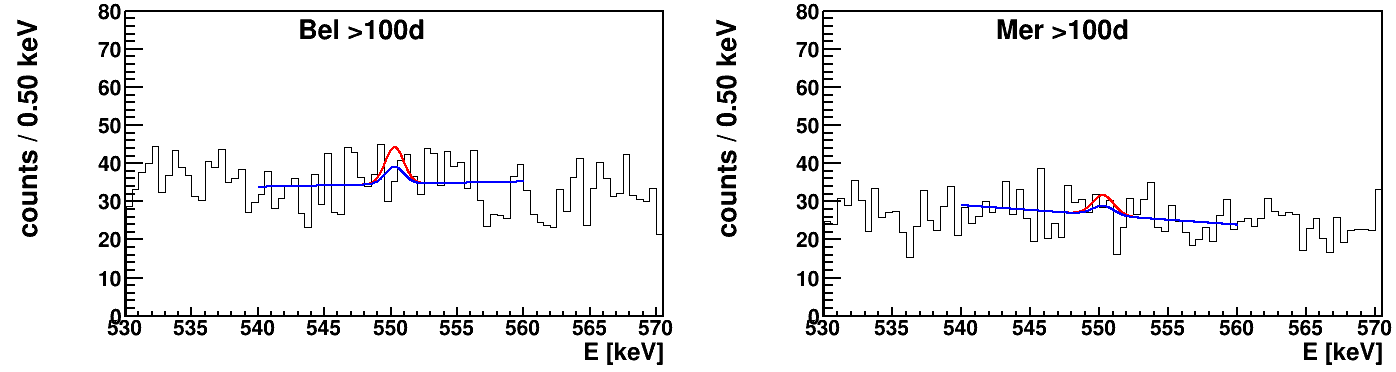}
\caption{Fit of the 550.3~keV \gline\ from \nuc{Gd}{152} $\alpha$ decay into the excited $2^+_1$ state (550.3~keV) of \nuc{Sm}{148}.}
  \label{pic:pdf_ROI_Composition_Alpha_Gd152_aES_GdSK_v5}
\end{figure*}

\begin{figure*}
  \centering
  \includegraphics[width=0.99\textwidth]{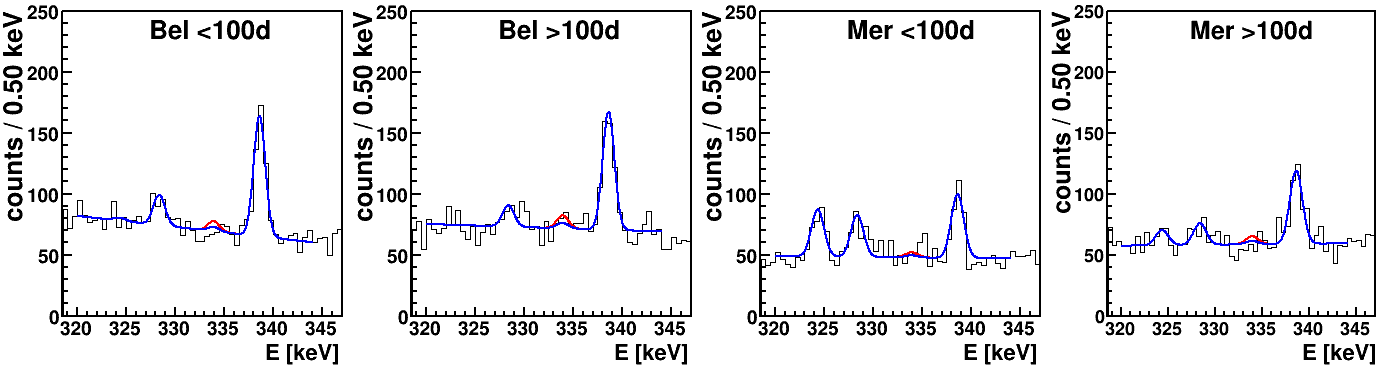}
\caption{Fit of the 333.9~keV \gline\ from \nuc{Gd}{154} $\alpha$ decay into the excited $2^+_1$ state (333.9~keV) of \nuc{Sm}{150}.}
  \label{pic:pdf_ROI_Composition_Alpha_Gd154_aES_GdSK_4DS_v5}
\end{figure*}

\begin{figure*}
  \centering

\begin{subfigure}{0.49\textwidth}  \includegraphics[width=0.99\textwidth]{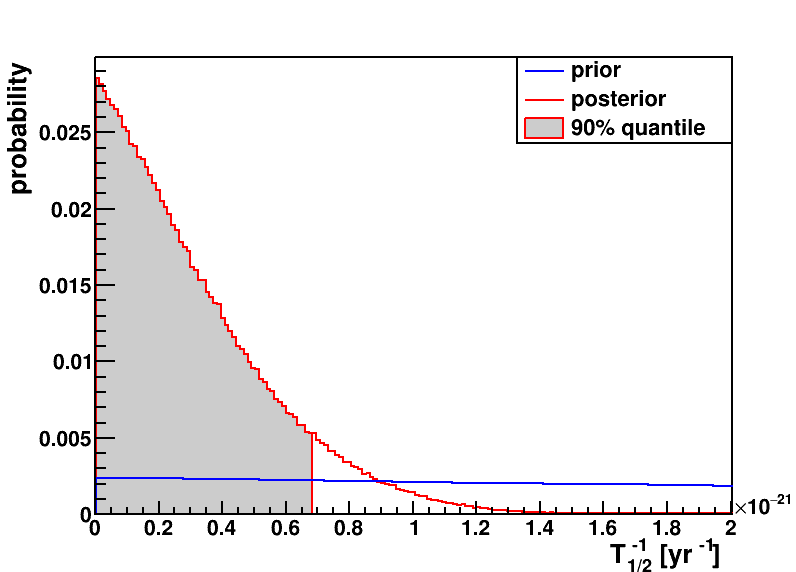} \caption{\nuc{Gd}{160} $0\nu/2\nu\beta\beta$ decay into the $2^+_2$ state}   
\end{subfigure}   
\begin{subfigure}{0.49\textwidth}   \includegraphics[width=0.99\textwidth]{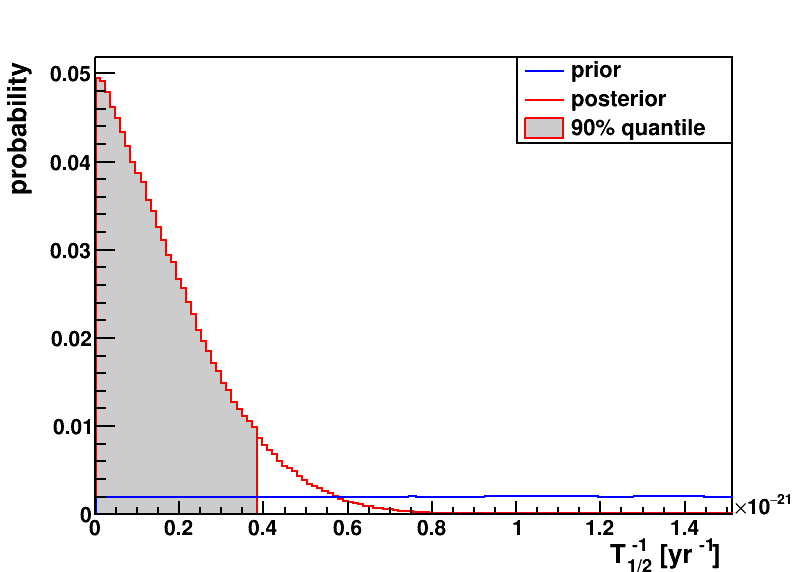} \caption{\nuc{Gd}{160} $0\nu/2\nu\beta\beta$ decay into the $0^+_1$ state} \end{subfigure}
\begin{subfigure}{0.49\textwidth}   \includegraphics[width=0.99\textwidth]{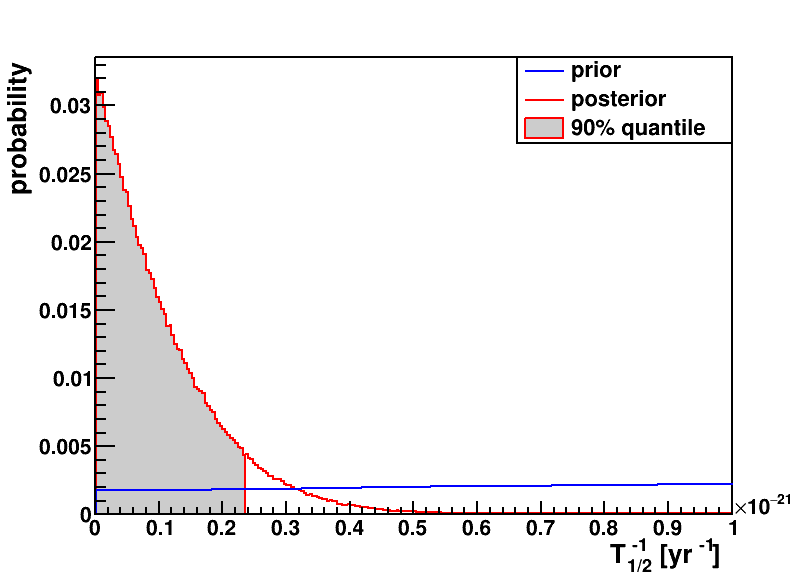} \caption{\nuc{Gd}{160} $0\nu/2\nu\beta\beta$ decay into the $0^+_2$ state} \end{subfigure}%
\begin{subfigure}{0.49\textwidth}  \includegraphics[width=0.99\textwidth]{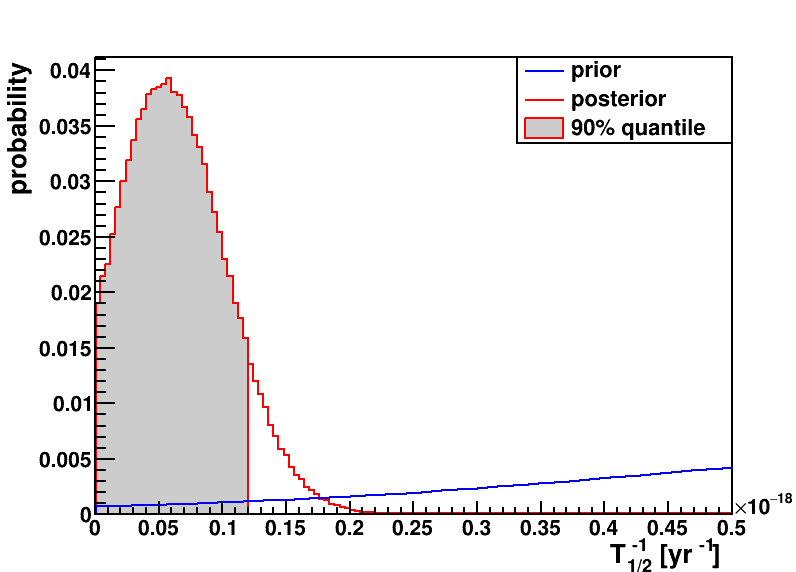} 
\caption{\nuc{Gd}{152} $\alpha$ decay into the $2^+_1$ state} \end{subfigure} 
\begin{subfigure}{0.49\textwidth}   \includegraphics[width=0.99\textwidth]{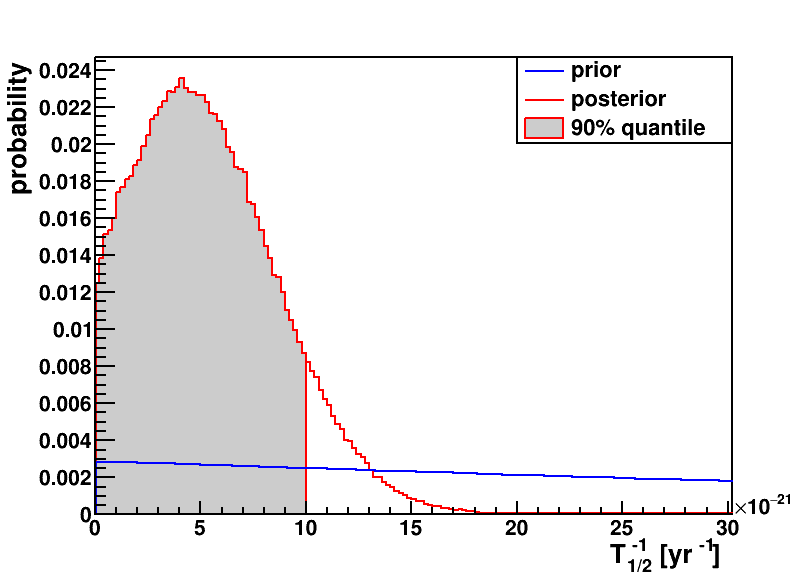} \caption{\nuc{Gd}{154} $\alpha$ decay into the $2^+_1$ state} \end{subfigure}%
  
\caption{Priors and posteriors of the Bayesian analysis for the inverse half-life variable. The shaded area show the 90\% quantile.}
  \label{pic:posteriors}
\end{figure*}

\end{document}